\RequirePackage{etex}
\documentclass{eptcs}
\providecommand{\event}{ICE 2018} 
\usepackage[T1]{fontenc}
\usepackage{amssymb}
\usepackage{amsmath}
\usepackage{mathrsfs}
\usepackage{amsthm}
\usepackage{proof}
\usepackage{stmaryrd}

\usepackage{graphicx}
\usepackage{listings}
\usepackage{bussproofs}
\usepackage{tikz}

\usetikzlibrary{arrows,shapes,trees,positioning,decorations,shapes}
\usetikzlibrary{decorations.markings,automata}

\newcommand{\RN}[1]{%
  \textup{\uppercase\expandafter{\romannumeral#1}}%
}

\usepackage{todonotes}
\newcommand{\ektodo}[1]{}
\newcommand{\ektodoin}[1]{}

\usepackage{ifthen}
\usepackage{xspace}
\usepackage{pdfpages}

\newtheoremstyle{mystyle}
{\topsep} 
{\topsep} 
{} 
{} 
{\bfseries} 
{\newline} 
{.5em} 
{} 
\theoremstyle{mystyle}

\newtheorem{lemma}{Lemma}

\let\temp\phi
\let\phi\varphi
\let\varphi\temp

\makeatletter
\newcommand{\xRightarrow}[2][]{\ext@arrow 0359\Rightarrowfill@{#1}{#2}}
\makeatother

\allowdisplaybreaks[1]

\colorlet{keywordcolor}{blue!50!black}
\colorlet{commentcolor}{green!60!black}
\colorlet{typecolor}{violet}
\newcommand{\sourcefont}{\ttfamily\small}
\newcommand{\commentfont}{\slshape\rmfamily\color{commentcolor}}

\lstdefinelanguage{ABS}{
        keywords={assert,this,new,data,type,def,case,of,local,class,interface,
        extends,implements,if,then,else,await,get,Fut,return,skip,while,foreach,module,
        import,export,from,to,suspend,delta,adds,modifies,removes,original,productline,
        features,core,corefeatures,optionalfeatures,after,when,product,hasAttribute,
        hasMethod,hasField,hasInterface,uses,root,extension,group,allof,oneof,require,
        stateupdate,objectupdate,classupdate,
        exclude,original,ifin,ifout,opt,null,
        newgroup,data,thiscomp,in,joins,leaves,subtypeOf,wait,acquire,except,as,component,Pre,Abs
        },
        keywordstyle=\color{keywordcolor}\bf\sffamily,
        morekeywords=[2]{Unit, Int, Bool, Rat, List, Set, Pair, Fut, Maybe, String, Triple, Either, Map},
        keywordstyle=[2]\color{typecolor},
        sensitive=true,
        comment=[l]{//},
        morecomment=[s]{/*}{*/},
        morestring=[b]"
}

\lstdefinelanguage[v9]{Java}[]{Java}{
        morekeywords={module,requires,provides,uses,with,to,exports}
}
\lstdefinelanguage[ContextJ]{Java}[]{Java}{
        morekeywords={layer,with,without,proceed,before,after}
}
\lstdefinelanguage[FOP]{Java}[]{Java}{
        morekeywords={refines,original,Super}
}
\lstdefinelanguage[JastAdd]{Java}[]{Java}{
        morekeywords={aspect,syn,inh,lazy}
}

\lstdefinestyle{code}{
        basicstyle=\sourcefont\upshape,
        keywordstyle=\color{keywordcolor}\bf\sffamily,
        commentstyle=\commentfont,
        columns=fullflexible,
        mathescape=true,
        escapechar={\#},
        keepspaces=true,
        showstringspaces=false,
        inputencoding=utf8,
        extendedchars,
        aboveskip=8pt, 
        numbers=left,
        stepnumber=1, 
        numberstyle=\ttfamily\scriptsize\color{gray},
        numbersep=4pt,
        xleftmargin=1.5em,
        xrightmargin=1.5em,
        framexleftmargin=1.2em,
        framexrightmargin=1em,
        framextopmargin=0.5ex,
        breaklines=true,
        breakindent=3pt,
}

\lstdefinestyle{abs}{
        style=code,
        language=ABS,
}
\lstdefinestyle{java}{
        style=code,
            language=Java
}
\lstdefinestyle{java9}{
        style=code,
            language=[v9]Java
}
\lstdefinestyle{aspectj}{
        style=code,
        language=[AspectJ]Java
}
\lstdefinestyle{jastadd}{
        style=code,
        language=[JastAdd]Java
}
\lstdefinestyle{contextj}{
        style=code,
        language=[ContextJ]Java
}
\lstdefinestyle{FOP}{
        style=code,
        language=[FOP]Java
}
\lstdefinestyle{scala}{
        style=code,
        language=Scala,
        morekeywords={self}
}

\newcommand{\code}[2][]{\lstinline[style=code,basicstyle=\ttfamily\upshape,#1]{#2}}

\newcommand{\abs}[2][]{\code[style=abs,#1]{#2}}

\lstnewenvironment{srccode}[1][]{
        \minipage{\linewidth}
        \lstset{style=code,
        backgroundcolor=\color{white},
        rulecolor=\color{gray!50},
        frame=tblr,
        captionpos=b,
        #1}
}{\endminipage}

\lstnewenvironment{abscode}[1][]{
        \minipage{1\linewidth}
        \lstset{style=abs,
        backgroundcolor=\color{white},
        rulecolor=\color{gray!50},
        frame=tblr,
        captionpos=b,
        #1}
}{\endminipage}

\lstnewenvironment{javacode}[1][]{
        \minipage{\linewidth}
        \lstset{style=java,
        backgroundcolor=\color{gray!5},
        rulecolor=\color{gray!50},
        frame=tblr,
        captionpos=b,
        #1}
}{\endminipage}

\lstnewenvironment{jastaddcode}[1][]{
        \minipage{\linewidth}
        \lstset{style=jastadd,
        backgroundcolor=\color{gray!5},
        rulecolor=\color{gray!50},
        frame=tblr,
        captionpos=b,
        #1}
}{\endminipage}

\newcommand{\xabs}[1]{\text{\abs{#1}}}

\newcommand{\EK}[1]{{\color{black}{#1}}}
\newcommand{\COMMENT}[1]{}

\title{Prototyping Formal System Models with Active Objects}

\author{Eduard Kamburjan
\institute{Technische Universit{\"a}t Darmstadt, Germany}
\email{kamburjan@cs.tu-darmstadt.de}
\and
Reiner H{\"a}hnle
\institute{Technische Universit{\"a}t Darmstadt, Germany}
\email{haehnle@cs.tu-darmstadt.de}
}
\def\titlerunning{Prototyping Formal System Models with Active Objects}
\def\authorrunning{E. Kamburjan \& R. H{\"a}hnle}
\begin{document}
\maketitle

\begin{abstract}
  We propose active object languages as a development tool for formal
  system models of distributed systems. Additionally to a formalization based on a
  term rewriting system, we use established Software Engineering
  concepts, including software product lines and object orientation
  that come with extensive tool support. We illustrate our modeling
  approach by prototyping a weak memory model.  The resulting
  executable model is modular and has clear interfaces between
  communicating participants through object-oriented modeling.
  Relaxations of the basic memory model are expressed as
  self-contained variants of a software product line. As a modeling
  language we use the formal active object language ABS which comes
  with an extensive tool set. This permits rapid formalization of core
  ideas, early validity checks in terms of formal invariant proofs, and
  debugging support by executing test runs. Hence, our approach
  supports the prototyping of formal system models with early
  feedback.
\end{abstract}

\section{Introduction}\label{sec:intro}

Formal methods provide formal frameworks for software and systems
development, including formally defined specification and programming
languages. Their aim is to support design and implementation of
engineering projects with high quality requirements, yet formal
notations themselves are developed without the very support they are
intended to provide. This is not simply an issue of productivity, but
of usability: one of the largest obstacles against the uptake of formal
methods is that they are expressed in---occasionally
dated---formalisms that are hard to communicate, to understand, and to
validate.

In this paper, we intend to show that recent progress in
\emph{software engineering}, including new structuring principles as
well as state-of-art tool support, can be beneficial in \emph{formal
  methods engineering} as well. Specifically, we use a modern active
object language instead of a term-rewriting system to formalize
semantics, we use software product lines (SPL)~\cite{variable} to organize and
maintain different variants of the formal model, and we use automated
theorem proving tools to provide early validity checks of the 
consistency of the formalized system.

We exemplify our approach with a representative case study in the area
of distributed systems, where we look at certain weak memory
models. The latter became very widespread, but their consequences are
notoriously hard to understand. Currently, weak memory models are
generally formulated as a term rewriting system for a small step
operational semantics (SOS) \cite{Boudol,Mantel}, as abstract
automata~\cite{Sarkar,xtso}, or as axiomatic/algebraic description of
traces~\cite{Axiom,Axiom2}.  The advantage of these formalisms is to
give the modeler freedom to adjust the formal semantics so it matches
the underlying intuition and to formulate properties of interest
without any restriction. One disadvantage is that the resulting formal
models are hard to understand for anyone who is not an expert in the
used formalism. Even for experts such formal models tend to be
hard to validate (i.e.\ debug), because limited tool support is
available. Available tools (e.g., rewriting engines
like Maude~\cite{maude}) lack the usability and modularity embodied in
more software-oriented tools. Finally, successful usage of currently
available tools requires that the formalization of the target system
is essentially finished, hence are not suitable for \emph{early}
prototyping \emph{during} formalization.

In the following we demonstrate that active object modeling is
adequate for the domain of weak memory models, and that it can offer
significant tool support during development, analysis, and
presentation of an operational semantics.  While weak memory models
can be regarded as representative and sufficiently complex, our
approach is in no way limited to this particular domain.

In our approach, an active object model is developed simultaneously
with an operational semantics. As a consequence, the modeler is able
to profit from debugging techniques, tool support, and best practices
of software engineering.  This includes support for modularization of
models, debugging by means of test runs, as well as automated proofs
of invariants.  We make use of four software engineering principles:
(1) \emph{modularization} with interfaces and modules, (2)
\emph{variant management} with software product lines, (3)
\emph{validation} by execution tests and formal verification, (4)
development by \emph{early prototyping} to obtain feedback and
experience with the product before creating its final version.  These
concepts permit to develop formal models faster, in an interactive
manner, and with higher confidence in their properties.

Our main contributions are: (1) a weak memory model, formalized in the
Abstract Behavioral Specification (ABS) language~\cite{ABS}; (2) a
discussion of the advantages of developing a prototype in an active
object language in parallel to the actual formalization.  The paper
is organized as follows: Section~\ref{sec:related} describes similar
approaches and compares to other methods for mechanizing formal
models. Section~\ref{sec:lang} introduces ABS.
Section~\ref{sec:weak} describes the formal model and its
implementation, Section~\ref{sec:sanity} describes the 
model validation, and Section~\ref{sec:dicuss} discusses the
advantages of our approach. Section~\ref{sec:concl} concludes.


\section{Related Work and Discussion of Tool-Based
  Approaches}\label{sec:related}

There is a recognized need for tool support to understand and analyze
relaxed memory models in mainstream programming languages. This led to
the implementation of simulators that are capable to explore the
configuration states generated from a given semantics, see Boudol
et al.~\cite{Boudol} in Java and Sarkar et al.~\cite{Sarkar} in
OCaml.  In either case the simulator is far larger and more complex
than the underlying semantics. The simulators are optimized for the
generation of configuration states.  Using general purpose languages
as state generators has two main downsides: (1) It obliterates the
differences between code expressing the model and code needed for a
framework to execute it, especially if the program has to be
optimized.  Reasoning about the model in terms of the simulation
program is hard, because it includes reasoning about the framework.
(2) The simulators are not intended to be used to \emph{communicate
  and reason} about the semantics, only to discover interesting
configurations, the modularization and structure they may have is not
transferable to the model.

Traditional tools used in work on formal semantics include term
rewriting engines such as Maude~\cite{maude}, CafeOBJ~\cite{cafe},
theorem provers like Isabelle/HOL~\cite{isabelle}, and model checkers
like SPIN/Promela~\cite{promela}. 
These approaches allow to state theorems about the model and often
have support for executability and modularization through namespaces.
However, they are mostly used as a more precise alternative to
pen-and-paper definitions and suffer from the similar downsides: (1) they
provide no additional modularity or interface abstractions---they do
not \emph{simplify} reasoning about a model, they merely
\emph{formalize} it. For example, Weber~\cite{weber} provides an
Isabelle formalization of the memory model of Mantel et
al.~\cite{Mantel} and finds notational errors that do not compromise
the theoretical results, but hamper comprehensibility.  These mistakes
were discovered, because the tool enforces syntax checks, not because
the formalization itself would provide a clearer structure.  (2)
traditional formalization tools are not designed for
\emph{prototyping} formal methods: i.e.\ to present and verify the
core ideas of a semantic \emph{before} fleshing it out.  In order to
validate formalization ideas by performing tool-based integrity checks
or to ensure that the system behaves as intended, the full model needs
to be formalized.  Some provers like Coq~\cite{coq} allow to assume
some properties without proving them for prototyping, but they already
require the structure of the final model to be fixed.  
Finally, (3)
current tools offer no specific support for the challenges of
distributed systems and they do not help to manage different variants
of a formal semantics. Libraries like K~\cite{kmaude} 
provide support for operational semantics, but to not address distributed systems or modularity.
This is a general drawback of general purpose tools: Prototyping relies on fast feedback cycles 
and general purpose approaches need libraries, which hamper analyzability or auxiliary code, which hampers clarity.

Active object languages like ABS~\cite{ABS} and Rebeca~\cite{rebeca}
try to combine the advantages of mainstream programming languages
(executability, fine-grained modularization) with the advantages of
languages with formal semantics (formal verification) and new ideas
from programming language research (e.g., variability management
through feature-based modeling~\cite{variable}) to simplify working
with concurrency. The enforced structure of actors and objects is too
restrictive to use them to analyze and express \emph{all} desirable
global properties, but their tool support and clarity predestines them
to prototype models of distributed systems: fast validation of core
ideas instead of full verification of the complete models.
Active object languages are used to formalize a vast range of
distributed systems: software systems~\cite{hadoop}, cyber-physical
systems~\cite{cyber1}, operational procedures~\cite{ftscs}, and
hardware~\cite{noc,noc2}.  We argue that distributed formal system
models can as well be modeled with active objects and that this
results in faster development and more efficient communication.


\section{Abstract Behavioral Specification}\label{sec:lang}

We provide a very short introduction to the \emph{Abstract Behavioral
  Specification} (ABS) language. For its formal semantics 
see Johnsen et al.~\cite{ABS}, 
Din et al.~\cite{ding}; for the formal semantics of product lines see
Muschevici et al.~\cite{absspl}, and for a tutorial see
H{\"a}hnle~\cite{tut}.

\subsection{Language}


ABS extends the actor~\cite{actor} concurrency model with
futures~\cite{future} and cooperative scheduling: Objects communicate
with each other only over \emph{asynchronous} method calls. Following
a method call, the caller receives a future as a handle for
identifying the starting process and continues execution. The callee
\emph{resolves} the future by storing its return value in it.  An
object may only switch the active process if the currently active
process explicitly releases control. This greatly simplifies the
concurrency model, as between the synchronization points a method can
be regarded as sequential.  A process releases control by either
terminating or suspending. The latter means either to wait for a
future to be resolved or for a condition to become true. Once a future
is resolved, any process that possesses a reference to it may read it,
because futures can be passed around. When a process attempts to read
from a non-resolved future, the whole object blocks until the future
is resolved.

ABS models can be compiled into Erlang, Haskell or Java code and be executed~\cite{ABS}.
Syntactically, ABS is close to Java and has similar concepts,
including interfaces and classes.  
However, classes have no static
fields and may not extend other classes. In addition, all objects are
strongly encapsulated: the only way to access their state is
via getter and setter methods they may declare.  
Modularization is supported by
syntax modules, akin to Haskell, that allow to import and export
classes, interfaces, and functions.  
The code in
Figure~\ref{fig:container} declares a module with a simple container
model.
We refrain from introducing the full ABS syntax, as most statements are
standard. We only describe the statements specific to the concurrency
model.
\begin{figure}[bht]
\begin{abscode}
module Container;
export *;
import Element from Element;
interface Container { 
  Unit setElement(Element e); 
  Element getElement(); 
}
class Container implements Container{
  Element contains = null;
  Unit setElement(Element e){ contains = e; }
  Element getElement(){ return contains; }
}
\end{abscode}
\caption{A simple \texttt{Container} in ABS.}  \label{fig:container}
\end{figure}
\begin{itemize}
\item To invoke (asynchronously) a method \abs{m} on the object stored
  in \abs{o}, the statement \abs{f = o!m(i)} is used. The value of
  \abs{i} is passed as the method parameter, and the resulting future
  is stored in \abs{f}.  The type of the future must match the return
  type of the method. 
\item There are two statements to synchronize with a future stored in
  \abs{f}: (1) \abs{f.get} attempts to synchronize on \abs{f} by
  reading its value. If \abs{f} is not resolved yet, then the process
  blocks until then: no other process may become active. (2) \abs{await f?} releases control of the
  object until \abs{f} is resolved, i.e., another process may become active.
\item To wait on a condition, the statement \abs{await b} releases
  control of the processor until the boolean expression evaluates to
  true.  The behavior is scheduler dependent, as the expression may
  evaluate to false again, if another process with such a side-effect
  is scheduled first.
\end{itemize}

We abbreviate the pattern \abs{Fut<T> f = o!m(); await f?; i = f.get;}
with \abs{i = await o!m();} and write \abs{foreach (i < E) \{ ... \}}
for \abs{Int i = 0; while (i < E) \{ ...; i = i +1; \}}.  If a class has
a method with the signature \abs{Unit run()}, then this method is
started automatically upon object creation.

ABS is object-oriented, but does not enforce to model everything as an object -- 
the enforced asynchronous communication leads to overhead for simple look-up operations. To omit
this, ABS uses \emph{Abstract Data Types} to abstract from data values
which have no internal state. 
%

\subsection{Product Lines}
\label{sec:product-lines}

ABS offers management of model variants via product lines~\cite{splbook}: A product
line describes different versions of a model that are obtained by
certain syntactic operations on a common core.  These syntactic
operations are called \emph{deltas} and they are able to add, as well
as replace classes, methods, and fields. The deltas are applied before
type checking.  When a method is replaced in a delta, then the new
version of the method may refer to the previous one with the keyword
\abs{original}.  The \abs{Notify} delta in upper part of
Figure~\ref{fig:line} modifies the \abs{setElement} method, such that
the element in the container is notified after being added and the
delta \abs{Queue} replaces the single contained element by a queue.

\begin{figure}[thb]
\begin{abscode}
delta Notify; 
  modifies class Container.Container {
    modifies Unit setElement(Element e) { 
      original();   
      await = e!observedBy(this); 
    }}
delta Queue; 
  modifies class Container.Container {
    removes Element contains;   
    adds List<Element> contains = Nil;
    modifies Unit setElement(Element e) { contains = Cons(e, contains); }
    modifies Element getElement() {
      Element e = value(contains); 
      contains = tail(contains); 
      return e; 
    }}
\end{abscode}

\begin{abscode}
productline ContainerElement; 
features QueueF, NotifyF;
delta Notify after Queue when NotifyF;
delta Queue when QueueF;
product NotifyProduct(NotifyF);
product FullProduct(QueueF, NotifyF);
\end{abscode}
\caption{Deltas and a product line in ABS.}  \label{fig:line}
\end{figure}

Product lines associate deltas with \emph{features} and describe
constraints among their application order. The \abs{Notify} delta must
be applied after \abs{Queue}, because only that version of
\abs{setElement} calls \abs{original}. This is achieved by the
\abs{after} directive in the lower part of Figure~\ref{fig:line}.

\subsection{Logic}

ABS offers \emph{invariant}-based reasoning to prove consistency and
safety properties of single objects. Safety and consistency are formulated as object invariants:
formulas in a first-order axiomatization of heap memory in a program
logic that must hold at every release point, i.e.\ at the end of each
method and whenever an \abs{await} statement is reached.  ABS methods
are integrated into the logic by \emph{modalities} over ABS
statements.  A calculus for showing validity of formulas in this logic
is built into the theorem prover KeY-ABS~\cite{keyabs}. It can verify
invariants semi-automatically by \emph{symbolic execution} of ABS
statements.  Symbolic execution can be used to compute formulas that
correspond to symbolic state transformers reflecting the state changes
caused by a given method.

The heap is axiomatized with functions to \textit{select} and
\textit{store} values to/from a reserved variable \textit{heap}.  To
express that a field \abs{i} is a list containing only positive values
one may use the following formula $\phi$:
\[
  \phi = \forall~\mathsf{Int}~k.~k\geq 0 \wedge k < \mathit{length}(\mathit{select}(\mathit{heap},\mathbf{self},i)) \rightarrow \mathit{select}(\mathit{heap},\mathbf{self},i)[k] > 0
\]

Strong encapsulation is reflected in the signature: each formula may
only reason about the fields of one class.  The following expresses
that $\phi$ is an invariant for a given piece of code: if it holds in
the beginning, then it holds after executing \abs{i = Cons(10,i);}:
\[\phi \rightarrow [\text{\abs{i = Cons(10,i);}}]\phi\]


\section{A Weak Memory Model}\label{sec:weak}

The memory systems used in modern hardware do not treat write and read accesses of processors as atomic.
Instead, memory accesses are stored in a queue before being executed. 
Before execution, they may be reordered and the execution of a read may read the requested value not from the memory, but from a not executed, but visible write access.
While this allows performance boosts, such \emph{weak memory models} are known to result in execution traces which are not reproducable by interleaving the executing threads~\cite{adve,lamport}

We present a simple weak memory model\footnote{The model is available at~\url{http://formbar.raillab.de/en/publications-and-tools/memory-model}} that is
able to simulate instruction reordering and write atomicity violation,
i.e.\ the two main principles of weak memory models~\cite{adve}.  Our
model is based on ideas taken from Boudol et al.~\cite{Boudol}, where
identifiers---similar to the futures in ABS---are used and from Mantel
et al.~\cite{Mantel}, where relaxation of the instruction order is
characterized by pair-wise comparisons on the queue of memory
accesses.  To keep this paper within reasonable length, we give no examples for some features of memory models, such as fences and visibility beyond read-own-write.

\subsection{Memory}
\label{sec:memory}

We start with the \abs{Memory} interface that models system
memory. Our core product variant is a conventional memory and does not
allow instruction reordering. Its central concept is the
\abs{Map<Location, Int> mem} field that maps location to values.  The
client interface is:

\noindent
\begin{abscode}
interface Memory{
  Fut<Int> read(Thread t, Location loc);
  Unit write(Thread t, Location loc, Int val);
  Int const(Int i);
}
\end{abscode}

The \abs{read}/\abs{write} methods model reading/writing a value
from/to the memory and \abs{const} returns a constant.  Internally,
memory access is managed by a \abs{list} of waiting accesses and each
call to \abs{read} or \abs{write} adds one access to that list.  A
single memory access has the following type:

\noindent
\begin{abscode}
data Access = Write(Thread tid, Location loc, Int value, Int id ) 
            | Read(Thread tid, Location loc, Int id);
\end{abscode}

Each memory \abs{Access} has a unique \abs{id} parameter. After adding
an \abs{Access}, the \abs{write} and \abs{read} methods terminate. The
\abs{read} method returns the future of a call to \abs{internalRead},
which is resolved once the access has been executed. I.e.\ it waits
until the \abs{id} of the corresponding \abs{Access} is added to the
\abs{done} set and then returns the read value, saved in the map
\abs{ret}.

\noindent
\begin{abscode}
Int read(Thread t, Location loc) {
  Int myId = counter; // global access counter
  list = appendright(list, Read(t, loc, myId));
  counter = counter + 1;
  return this!internalRead(myId); 
}
Int internalRead(Int myId) {     
  await contains(done, myId); 
  return lookupUnsafe(ret, myId);//Note: value is not read from mem
}
\end{abscode}

The memory is modeled with a loop that first waits until the
\abs{list} of scheduled \abs{Access} items is not empty, see
Figure~\ref{fig:big}. Then it invokes a \abs{strategy} which returns
the positions of those accesses that can be safely executed next.
Afterwards, it gets and removes the scheduled access from the list,
and executes its effect. The value \abs{snd(pp)} is obtained from a
call to \abs{getValueFor} and used in the read. In case of a write,
the given value is written to memory.  The \abs{strategy} method
implements the procedure defined in Mantel et~al.~\cite{Mantel}: At
each position $i$ of \abs{list}, the access at position $i$ is
compared to all accesses on positions $j<i$. The \abs{maySwap} method
takes two accesses and decides whether the second access can be executed before the first one.
In the core product, i.e. under sequential consistency, it returns \abs{False} if and only if the two
accesses are from the same thread. The \abs{getValueFor} method simply
reads from memory.

\begin{figure}[tbh]
\begin{abscode}
Unit run() {
  while (True) {
    await list != Nil; 
    List<Int> accList = this.strategy();
    Pair<Access, Int> pp = this.getAccess(cextPos(accList));  // invokes getValueFor
    case fst(pp) {
      Write(t, loc, val, id) => 
        { done = insertElement(done, id); mem = put(mem, loc, val); }
      Read(t, loc, id)       => 
        { done = insertElement(done, id); ret = put(ret, id, snd(pp)); }
    }       	   
  }
}
List<Int> strategy() {
  List<Int> allowed = Nil;
  foreach (i < length(list)) { // prettified
    Bool add = True;
    foreach (j < i) { // prettified
      Bool b = this.maySwap(nth(list,j), nth(list,i));
      if (!b) { add = False; }
    }
    if (add) { allowed = Cons(i, allowed); }
  }
  return allowed;
}
Int getValueFor(Thread tid, Location loc, Int pos) { 
  return lookupUnsafe(mem, loc); 
}
\end{abscode}
\caption{The next access loop and the access scheduling strategy.}
\label{fig:big}
\end{figure}



\subsection{Threads}

We aim to describe a memory model, not a whole programming language,
hence threads are series of memory invocations. We do not model
control flow or make assumptions about specific properties of the
local store.  A memory invocation can have one of the following forms:

\begin{enumerate}
\item \abs{await mem!write(this,Location(name),val);} models a write
  access. Observe that after \abs{await} execution continues once the
  \abs{Access} is added to the \abs{list}, not when the \abs{Access}
  is executed.
\item \abs{reg = await mem!read(this,Location(name));} models a read
  access that reads the future into a local location \abs{reg}.
\item \abs{reg.get;} models an access where the value is retrieved,
  i.e.\ a computation.
\end{enumerate}
The \abs{await} statements enforce a FIFO treatement of \emph{adding} memory accesses to the queue, from the point of view of a single process.
The future returned by \abs{read} is then used to synchronize on the \emph{execution} of the added memory access.
This is needed for two reasons: (1) Memory systems are synchronous by their nature and it is necessary to enforce synchronicity \emph{at this point}, despite using a asynchronous language.
(2) Additionally to asynchronous communication, ABS does not enforce FIFO scheduling, i.e., the callee object is not required to process the calls in the order they arrive.

\subsection{Instruction Reordering}

To include instruction reordering into our model, we define the
predicates for program order relaxation described by Mantel et\ al.\
in terms of deltas (Section~\ref{sec:product-lines}). Each relaxation
predicate adds a condition in the \abs{maySwap} method.  For example,
to permit write-read reordering, one applies the following delta:

\noindent
\begin{abscode}
delta WRDelta; 
modifies class Mem.Memory {
  modifies Bool maySwap(Access a, Access b) {
    Bool ret = False;
    case a {
      Read(_,_,_)       => { ret = original(a,b); }
      Write(_,loca,_,_) => case b { 
        Read(_,locb,_) => { last = original(a,b); #\label{line:write}#
                            ret  = (last || loca != locb); }#\label{line:write2}#
        Write(_,_,_,_) => { ret  = original(a,b); }}
    }
    return ret;
  }}
\end{abscode}

This method checks whether the arguments are write and read accesses
on different locations. The default is a call to \abs{original}.  The
first access for each thread may be executed first, because the core
product models the condition that accesses from different threads can
always be reordered and \abs{original} is called in each case.  The
remaining three reorderings are defined in an analogous manner.

\subsection{Write-Atomicity Violation}

To allow read-own-write, the \abs{getValueFor} method must be changed:
When executing a read at position \abs{pos}, it checks
whether there is a write \emph{from the same thread} in the access
list before \abs{pos}.  This corresponds to local buffers for each
thread. We refrain from modeling more fine-grained visibility for 
read-others-early for presentation's sake.

\noindent
\begin{abscode}
delta ReadEarlyDelta; 
modifies class Mem.Memory {
  modifies Int getValueFor(Thread tid, Location loc, Int pos) {
    return case getWriteValueFor(slice(list, 0, pos-1), loc, tid) {
              Just(val) => val;
              Nothing   => original(loc, pos);
           };
  }}
\end{abscode}

Method \abs{maySwap} must now allow write-read reordering
when both arguments access the same location. An additional delta \abs{WROwnDelta} does this. 
It is identical to \abs{WRDelta}, except that the statements in 
Lines~\ref{line:write}-\ref{line:write2} are replaced by 
\abs{if (loca==locb) ret = True; else ret  = original(a,b);}. 

Different memory models are modeled as product variants. Each memory
model is obtained from a set of conditions for instruction reordering
and the product corresponding to that model results from the
application of appropriate deltas.  The resulting product line is in
Figure~\ref{fig:pl}.
%

\begin{figure}[tbh]
\begin{abscode}
productline Memory;
features WWFeature, WRFeature, ReadEarlyFeature;
delta WWDelta when WWFeature;
delta WRDelta after WROwnDelta when WRFeature;
delta ReadEarlyDelta when ReadEarlyFeature;
delta WROwnDelta when ReadEarlyFeature;
product TSO (WRFeature, ReadEarlyFeature);
product PSO (WRFeature, WWFeature, ReadEarlyFeature);
product IBM370 (WRFeature); 
\end{abscode}
  \caption{Declaration of the weak memory product line, following the terminology of Mantel et al.~\cite{Mantel}.}
  \label{fig:pl}
\end{figure}


\section{Model Validation}\label{sec:sanity}

The capability to \emph{validate} design decisions made during
formalization at an early stage can save a lot of work and helps to
improve trust.  We exemplify the validation with our model by three
basic integrity checks: (1) the weakened model can produce traces that
are only obtainable in a relaxed, not in a conventional, memory model;
(2) the memory model cannot deadlock, provided that the threads using
it do not deadlock, and (3) the modeling of \abs{id}s is consistent.

\subsection{Instruction Reordering}\label{sec:litmus}

A classical litmus test, i.e., a test to determine whether a weak memory model is able to reproduce basic effects of non-atomic memory accesses, taken from Boudol et al.~\cite{Boudol}, is
the following simplification of Dekker's algorithm:
\begin{center}
\abs{v = 1; println("tr1: "+w)}\qquad run in parallel with\qquad \abs{w = 1; println("tr2: "+v)} 
\end{center}

This program can result in a trace where $\mathtt{v} = \mathtt{w} = 0$
holds in the final state only under a weak memory model with
write-read relaxations and relaxed write atomicity,
e.g. IBM370 or TSO~\cite{adve}.  This behavior is reproducible in our model. The
code in Figure~\ref{fig:ex00} models the memory accesses of the
program above.  In the product \abs{TSO}, the reordering can be
observed by running the ABS model:

\begin{verbatim}
tr1: 0  
tr2: 0
\end{verbatim}

\noindent To enforce this behavior, one may, for example, let the memory system
start after 3s and force that reads are executed before writes
whenever possible.\footnote{Product \abs{TSODemo}, resp. \abs{IBMDemo} in the downloadable
  model.}
 In the core product, this behavior is not observable.  This
 correlates with the fact that this behavior is not observable in architectures
 with sequential consistency.

\begin{figure}[tbh]
\begin{minipage}{0.49\textwidth}
\begin{abscode}
class T1(Memory mem) implements T {
 Fut<Int> rg;
 Unit run() {
  rg = mem.const(0);
  await mem!write(this, Loc("v"), 1);  
  rg = await mem!read(this,Loc("w"));    
  println("tr1: "+toString(rg.get));
 }
}
\end{abscode}
\end{minipage}
\begin{minipage}{0.49\textwidth}
\begin{abscode}
class T2(Memory mem) implements T {
 Fut<Int> rg;
 Unit run() {
  rg = mem.const(0);
  await mem!write(this, Loc("w"), 1);  
  rg = await mem!read(this,Loc("v"));    
  println("tr2: "+toString(rg.get));
 }
}
\end{abscode}
\end{minipage}
\caption{Encoding of the example as a sequence of memory accesses.}
\label{fig:ex00}
\end{figure}

\subsection{Local Deadlock}

A highly desirable property of any memory model is that it should not
deadlock, provided that the whole program does not deadlock. The
notion of deadlock in ABS is a \emph{circular} dependency between
multiple processes~\cite{dead}, where a process $p_1$ is said to
\emph{depend on} a process $p_2$, if $p_1$ is halting at a \abs{f.get}
or \abs{await} statement and execution of $p_2$ would allow $p_1$ to
continue. We refrain from introducing the full semantics of deadlocks
and execution, which are non-trivial in presence of condition synchronization~\cite{avocs}, but give a proof sketch for the following lemma:
\begin{lemma}\label{lem:lem}
  There is no reachable state in any valid ABS program using the class
  \abs{Mem} that contains a deadlock consisting only of processes
  running on an instance of \abs{Mem}.
\vspace{-3mm}
\begin{proof}[Proof sketch]
  First, we observe that there are no \abs{f.get} statements and only
  two \abs{await} statements, both with boolean guards: one in the
  \abs{run} method, one in \abs{internalRead}. Hence, every
  deadlock can only involve processes running \abs{run} and
  \abs{internalRead}.  Second, after the \abs{await} statement of
  \abs{internalRead}, the method has no side effects, therefore, no
  other process can depend on a process running \abs{internalRead}. So
  any deadlock can only involve processes running \abs{run}. But there
  is no call of \abs{run}, so there is only one process running
  \abs{run}.  There can be no deadlock, because for this at least two
  processes are needed.\qedhere
\end{proof}
\end{lemma}
A fully automatic deadlock-detection tool for
ABS\footnote{\url{http://formbar.raillab.de/deadlock}} is available, but it
requires a complete system, not merely standalone classes. The code in
Figure~\ref{fig:ex00} is analyzed as deadlock-free in the core
product, with one false positive \emph{live}lock risk due to the
non-terminating loop. Further deadlocks are not possible, because our
thread model so far contains only memory accesses, but no
synchronization statements.

\subsection{Invariants}
\label{sec:invariants}

We show an invariant about internal consistency of our model:
there are no two memory accesses with the same \abs{id}.
%
This is a direct consequence of another invariant: the value of the
\abs{counter} field is always larger than the highest \abs{id}
occurring in the \abs{list}.  We formulate these two invariants in
first-order logic:
\begin{align}
&\forall~ \mathbb{N}~ i,j.~i \neq j \wedge i < \mathsf{length(\xabs{list})} \wedge j < \mathsf{length(\xabs{list})} \rightarrow \mathsf{id}(\xabs{list}[i]) \neq \mathsf{id}(\xabs{list}[j])\\
&\forall~ \mathbb{N}~ i.~ i < \mathsf{length(\xabs{list})} \rightarrow \mathsf{id}(\xabs{list}[i]) < \xabs{counter}
\end{align}

These invariants can be shown using the KeY-ABS prover \emph{fully
  automatically}\footnote{Invariants and KeY-ABS are included with the
  downloadable model. We had to formalize the theory of \abs{Access}
  and \abs{List}.}  for the \abs{read} and \abs{write} methods. The
other methods trivially preserve these properties, as they do not add
elements to \abs{list} nor modify \abs{counter}. This can be as well
verified with KeY-ABS.  The proofs for \abs{read} and \abs{write} have
640 and 851 steps, respectively, and take a few seconds.


\section{Discussion}\label{sec:dicuss}

Our model conceptually follows the one by Mantel et al.~\cite{Mantel},
so we mostly compare it to theirs.  Their main goal is to establish
non-interference results for weak memory models, while ours is to
produce a modular, easy-to-follow, and analyzable
formalization. Therefore, both approaches are not exclusive, but
complementary: Rewriting systems can establish more complex
properties, while active object languages help during formalization.

\paragraph{Modularity.}
We are able to provide a small and clear interface to program memory,
see Section~\ref{sec:memory}. This is difficult to achieve in term
rewriting system-based models which do not syntactically distinguish
rules for executing language and memory statements.  In the model
of~\cite{Mantel}, for example, memory access execution is spread among
three different predicates in differing rule premises.  Moreover, the
parameters of predicates that model \emph{different} aspects are not
independent. For example, the predicate describing that an access at
position $i$ can be executed takes $i$ as a parameter, while the
predicate describing the read value takes the list $[0,\dots,i-1]$ as
parameter.

The ABS model we created is in very close correspondence to the
modeled concepts. Once the ABS model is in place, these \emph{can be
  carried over} also to a term rewriting semantics: methods may be
mapped to functions or predicates, interfaces to predicate signatures,
fields to configuration elements, etc. The three different rule
premises for executing an access mentioned above, for example, can be
encapsulated into a single function that corresponds to one loop
iteration of the \abs{run} method.  The ABS model helps to find such
abstractions by enforcing a process-oriented view.

A small and clear interface reduces the risk of clerical errors in the
formalization. It is justified by the fact that threads and system
memory also have a clear interface in hardware.  A concise interface
also helps to transfer the memory model to another language. In both
models \cite{Boudol,Mantel} the memory access aspects are polluted
with specific constructs from the language in which the memory model
resides: registers, special representations of identifiers, etc. An
active object model could have helped to encapsulate the
language-independent parts better.
%
%
The ABS model makes no assumption about the implementation language,
as it uses futures, not specific concepts that need to be modeled on
top, yet it shows where and how modularity can be achieved.

\paragraph{Product Variant Management.}
Our model supports the generation of multiple variants. The
modifications that produce these variants are \emph{not visible} in
the core product and hence does not pollute it. The standard approach
in rewriting systems to support different variants is to parameterize
it with underspecified functions or predicates that can be constrained
by different axioms. In contrast to this, the product line approach
expresses constraints by invariants and declaration of software
product lines. This keeps the analysis simple for the core product,
i.e.\ the basic case, as default values for parameters are
unnecessary. It alleviates also analyzing the differences between
different products, because the modifications are encapsulated in
deltas and not scattered around.

Variability management is orthogonal to modularity management, which is offered by , e.g., mixins or inheritance.
Product lines allow to encapsulate changes to more than one class and even single deltas may define operations on multiple classes.
Furthermore, variability is not managed inside the type system -- after variant generation, the program contains no product lines constructs anymore and
single variants can be analyzed or compiled with the tools already available for non-variant programs.

A discussion on the difference between intra-variant and inter-variant code reuse can be found in~\cite{damitrait}, a discussion on the relation of product lines to other concepts in~\cite{Schaefer2012}.

\paragraph{Debugging and Validation.}
In this section we discuss our approach to validation from Section~\ref{sec:sanity}.
The deadlock check (Lemma~\ref{lem:lem}) makes not use of a deadlock
analysis tool, but of the \emph{concept of a deadlock}.  In ABS, the
concept of deadlock is non-trivial (due to await on Boolean
conditions), but natural, because ABS has notions of progress and unit
of computation (process).  These notions do not exist in term
rewriting systems, making it hard to even state Lemma~\ref{lem:lem}:
rewrite rules for operational semantics often have multiple
preconditions and it is not obvious which of them may block.

The execution witnesses from the litmus test on instruction reordering
in Section~\ref{sec:litmus} are an example of minimal requirements one
may want to ensure. 
o
While we did not provide a proof that the core product does not allow
certain states, it would have been possible to formulate suitable
invariant, similar as in Section~\ref{sec:invariants}, even though it
might not be fully automatic. Active object languages like
Rebeca~\cite{rebeca} that offer model checking support would deliver a
\emph{fully automatic} proof.

In contrast to mainstream general purpose languages, ABS has a fully
formal semantics, allows to state formal invariants, and comes
with a verification tool that is able to verify such invariants.  The
invariants in Section~\ref{sec:invariants} were shown
automatically. Proofs with small amount of interaction might also be
acceptable, but the amount of time to invest into formal
validation depends on the modeled system.

In general, we conjecture that invariant proofs with model checking is
adequate for consistency conditions, but too limited for complex
properties such as non-interference, in particular, when
\emph{multiple} memory models are involved.  Thus, validation can be
used to ensure that properties hold during development, but it cannot
replace a full formalization.

\paragraph{Testing.}
Executability of ABS models and the high automatization of KeY (for simple invariants) and other static analysis,
allows to intergrate automatic testing into the prototyping apporach. 
One can see the example given in Section~\ref{sec:litmus} as a test, which can be part of a regression test suite that 
is used to ensure that the introduction of new feature to the model does not break old sanity checks.
The invariants in Section~\ref{sec:invariants} can also be used as regression tests, while the non-automatic Lemma~\ref{lem:lem} does not.
However, some fixed programs can be used as input to the static deadlock analysis to complete the test suite.

We point out that ABS supports unit testing and unit test generation~\cite{apet}, but we did not use this capability in this work.

\paragraph{Prototyping.}
ABS is suitable for quick development: the first model presented
above, including the first two validity checks were completed in under.\
three modeling hours by one person familiar with ABS.
The first model implementing the concepts of \cite{Mantel} including
all three validity checks was ready after one more work day (by the
same modeler).  Further changes for the presentation, bug fixing, and
the proof of the invariant took three more hours. Altogether, the
whole development presented in this paper took less than two work
days, but, of course, we had read and benefited from
\cite{Boudol,Mantel}.

The modeler made the experience that it was easier to explain the
ABS model than the rewrite system-based model, even to researchers
with experience in rewriting systems. Interestingly, one colleague
found several bugs in a preliminary version of our model related to
reordering, despite being only provided the same code segments
that are presented in this paper, and not being familiar with weak
memory models. \EK{We assume that programming-language-like modeling languages are more accessible and thus a better formalism to expose computer scientists and engineers to such models.
This is, however, only anecdotical evidence, as a proper user study is out of the scope of this work.}


\paragraph{Generalization and Limitations.}
In this paper we chose the ABS language, mainly because of the
similarities between futures and the identifiers of Boudol et
al.~\cite{Boudol}, as well as to perform a validity check through a
consistency invariant.  Other active object languages would also be
possible, for example, Rebeca~\cite{rebeca} that comes with a model
checker.

Compared to the approaches discussed in Section~\ref{sec:related}, 
Active Objects are less flexible. While they provide a framework for \emph{computation with asynchronous communication},
they do not allow to deviate from this principles.
E.g., active object languages enforce
strong encapsulation---if the modeled domain does not have strong
encapsulation of its elements, then other approaches to prototype
distributed systems might be a better fit.

The choice of modeling language is influenced by which kind of
validity checks are desired and by the desire to have a good match of
concepts between the domain and the model.  We suspect that active
object languages are a less good choice for sequential systems.

From a process perspective, the use of static analysis tools, makes the modeler more dependent on the developers.
E.g., 
if the deadlock analysis tool is not precise enough, one must extend the existing tool, which creates overhead for getting acquainted with the code base etc. The sheer presence of a static analysis does not ensure that its results are useful for prototyping
Similarely, 
there is no non-interference analysis tool for ABS, which makes it hard to reason about security or recreate the main results from Mantel et al.

Finally, in a term rewriting approach the modeler can change the model according to his proof strategy, while the semantics of the active object language is fixed. 
This may result in a demand for sanity checks, which are hard to express in active objects, even on the prototyping stage of development.



\section{Conclusion}\label{sec:concl}

We demonstrated that a model in an active object language results in a
concise and easy-to-comprehend formal model of system memory.  While
such a formalization cannot replace a full-fledged formalization as a
rewriting system or an abstract machine, it can help during modeling
core features, as well to communicate and to structure the essential
ideas.  We showed that \emph{message-based} modeling with Active Objects is to some extent natural to
model \emph{shared memory-based} concurrency. Based on
this observation, we speculate that by modeling memory explicitly, the
tools developed for actors and active objects are applicable for
shared memory scenarios, including high performance computing.

\paragraph{Future Work.}
We intend to model in ABS the approach of Bijo et al.~\cite{Bijo} for
cache architectures and extend it with a write buffer to achieve a
weak memory model.  We aim to use the ABS tools for resource
analysis~\cite{costabs} on different caching strategies to demonstrate
that the analytic possibilities of active objects can go beyond what
is possible in conventional non-trivial formal models.  When doing
this, we plan to develop the ABS model and the term rewriting
extension of \cite{Bijo} in parallel, to test the prototyping approach
advocated here.

The main theorem of Mantel et al. that non-interference under
one memory model does not imply non-interference under another memory
model is proven by providing a program and a distinguishing condition
for each pair of memory models, such that the program fulfills the
condition under one memory model, but not under the other.  As 
the provided programs contain no loops, we assume that a model checking
approach with Rebeca which would simplify the current proof is feasible.




\bibliographystyle{eptcs}
\bibliography{bibliography}

\begin{thebibliography}{10}
\providecommand{\bibitemdeclare}[2]{}
\providecommand{\surnamestart}{}
\providecommand{\surnameend}{}
\providecommand{\urlprefix}{Available at }
\providecommand{\url}[1]{\texttt{#1}}
\providecommand{\href}[2]{\texttt{#2}}
\providecommand{\urlalt}[2]{\href{#1}{#2}}
\providecommand{\doi}[1]{doi:\urlalt{http://dx.doi.org/#1}{#1}}
\providecommand{\bibinfo}[2]{#2}

\bibitemdeclare{article}{adve}
\bibitem{adve}
\bibinfo{author}{Sarita~V. \surnamestart Adve\surnameend} \&
  \bibinfo{author}{Kourosh \surnamestart Gharachorloo\surnameend}
  (\bibinfo{year}{1996}): \emph{\bibinfo{title}{Shared Memory Consistency
  Models: {A} Tutorial}}.
\newblock {\sl \bibinfo{journal}{{IEEE} Computer}}
  \bibinfo{volume}{29}(\bibinfo{number}{12}), pp. \bibinfo{pages}{66--76},
  \doi{10.1109/2.546611}.

\bibitemdeclare{inproceedings}{costabs}
\bibitem{costabs}
\bibinfo{author}{Elvira \surnamestart Albert\surnameend}, \bibinfo{author}{Puri
  \surnamestart Arenas\surnameend}, \bibinfo{author}{Samir \surnamestart
  Genaim\surnameend}, \bibinfo{author}{Miguel \surnamestart
  G\'{o}mez-Zamalloa\surnameend} \& \bibinfo{author}{Germ\'{a}n \surnamestart
  Puebla\surnameend} (\bibinfo{year}{2012}): \emph{\bibinfo{title}{{COSTABS}: A
  Cost and Termination Analyzer for {ABS}}}.
\newblock In: {\sl \bibinfo{booktitle}{Proc.\ ACM SIGPLAN Workshop on Partial
  Evaluation and Program Manipulation}}, \bibinfo{publisher}{ACM}, pp.
  \bibinfo{pages}{151--154}, \doi{10.1145/2103746.2103774}.

\bibitemdeclare{inproceedings}{apet}
\bibitem{apet}
\bibinfo{author}{Elvira \surnamestart Albert\surnameend}, \bibinfo{author}{Puri
  \surnamestart Arenas\surnameend}, \bibinfo{author}{Miguel \surnamestart
  G{\'{o}}mez{-}Zamalloa\surnameend} \& \bibinfo{author}{Peter Y.~H.
  \surnamestart Wong\surnameend} (\bibinfo{year}{2013}):
  \emph{\bibinfo{title}{aPET: a test case generation tool for concurrent
  objects}}.
\newblock In \bibinfo{editor}{Bertrand \surnamestart Meyer\surnameend},
  \bibinfo{editor}{Luciano \surnamestart Baresi\surnameend} \&
  \bibinfo{editor}{Mira \surnamestart Mezini\surnameend}, editors: {\sl
  \bibinfo{booktitle}{Joint Meeting of the European Software Engineering
  Conference and the {ACM} {SIGSOFT} Symposium on the Foundations of Software
  Engineering, ESEC/FSE'13}}, \bibinfo{publisher}{{ACM}}, pp.
  \bibinfo{pages}{595--598}, \doi{10.1145/2491411.2494590}.

\bibitemdeclare{article}{future}
\bibitem{future}
\bibinfo{author}{Henry~C. \surnamestart Baker\surnameend, Jr.} \&
  \bibinfo{author}{Carl \surnamestart Hewitt\surnameend}
  (\bibinfo{year}{1977}): \emph{\bibinfo{title}{The Incremental Garbage
  Collection of Processes}}.
\newblock {\sl \bibinfo{journal}{SIGART Bull.}} (\bibinfo{number}{64}), pp.
  \bibinfo{pages}{55--59}, \doi{10.1145/872736.806932}.

\bibitemdeclare{inproceedings}{Bijo}
\bibitem{Bijo}
\bibinfo{author}{Shiji \surnamestart Bijo\surnameend},
  \bibinfo{author}{Einar~Broch \surnamestart Johnsen\surnameend},
  \bibinfo{author}{Ka~I. \surnamestart Pun\surnameend} \&
  \bibinfo{author}{S.~Lizeth~Tapia \surnamestart Tarifa\surnameend}
  (\bibinfo{year}{2016}): \emph{\bibinfo{title}{An Operational Semantics of
  Cache Coherent Multicore Architectures}}.
\newblock In: {\sl \bibinfo{booktitle}{Proceedings of the 31st Annual ACM
  Symposium on Applied Computing}}, \bibinfo{series}{SAC '16},
  \bibinfo{publisher}{ACM}, pp. \bibinfo{pages}{1219--1224},
  \doi{10.1145/2851613.2851718}.

\bibitemdeclare{inproceedings}{Boudol}
\bibitem{Boudol}
\bibinfo{author}{G{\'{e}}rard \surnamestart Boudol\surnameend},
  \bibinfo{author}{Gustavo \surnamestart Petri\surnameend} \&
  \bibinfo{author}{Bernard~P. \surnamestart Serpette\surnameend}
  (\bibinfo{year}{2012}): \emph{\bibinfo{title}{Relaxed Operational Semantics
  of Concurrent Programming Languages}}.
\newblock In \bibinfo{editor}{Bas \surnamestart Luttik\surnameend} \&
  \bibinfo{editor}{Michel~A. \surnamestart Reniers\surnameend}, editors: {\sl
  \bibinfo{booktitle}{{EXPRESS/SOS}, Proc.}}, {\sl
  \bibinfo{series}{{EPTCS}}}~\bibinfo{volume}{89}, pp. \bibinfo{pages}{19--33},
  \doi{10.4204/EPTCS.89.3}.

\bibitemdeclare{inproceedings}{Axiom}
\bibitem{Axiom}
\bibinfo{author}{Sebastian \surnamestart Burckhardt\surnameend},
  \bibinfo{author}{Madanlal \surnamestart Musuvathi\surnameend} \&
  \bibinfo{author}{Vasu \surnamestart Singh\surnameend} (\bibinfo{year}{2010}):
  \emph{\bibinfo{title}{Verifying Local Transformations on Relaxed Memory
  Models}}.
\newblock In \bibinfo{editor}{Rajiv \surnamestart Gupta\surnameend}, editor:
  {\sl \bibinfo{booktitle}{Compiler Construction: 19th Intl.\ Conf.\, CC}},
  \bibinfo{publisher}{Springer}, pp. \bibinfo{pages}{104--123},
  \doi{10.1007/978-3-642-11970-5\_7}.

\bibitemdeclare{book}{maude}
\bibitem{maude}
\bibinfo{author}{Manuel \surnamestart Clavel\surnameend},
  \bibinfo{author}{Francisco \surnamestart Dur\'{a}n\surnameend},
  \bibinfo{author}{Steven \surnamestart Eker\surnameend},
  \bibinfo{author}{Patrick \surnamestart Lincoln\surnameend},
  \bibinfo{author}{Narciso \surnamestart Mart\'{\i}-Oliet\surnameend},
  \bibinfo{author}{Jos{\'e} \surnamestart Meseguer\surnameend} \&
  \bibinfo{author}{Carolyn \surnamestart Talcott\surnameend}
  (\bibinfo{year}{2007}): \emph{\bibinfo{title}{All About Maude - a
  High-performance Logical Framework: How to Specify, Program and Verify
  Systems in Rewriting Logic}}.
\newblock \bibinfo{publisher}{Springer-Verlag},
  \doi{10.1007/978-3-540-71999-1}.

\bibitemdeclare{inproceedings}{damitrait}
\bibitem{damitrait}
\bibinfo{author}{Ferruccio \surnamestart Damiani\surnameend},
  \bibinfo{author}{Reiner \surnamestart H{\"{a}}hnle\surnameend},
  \bibinfo{author}{Eduard \surnamestart Kamburjan\surnameend} \&
  \bibinfo{author}{Michael \surnamestart Lienhardt\surnameend}
  (\bibinfo{year}{2017}): \emph{\bibinfo{title}{A Unified and Formal
  Programming Model for Deltas and Traits}}.
\newblock In \bibinfo{editor}{Marieke \surnamestart Huisman\surnameend} \&
  \bibinfo{editor}{Julia \surnamestart Rubin\surnameend}, editors: {\sl
  \bibinfo{booktitle}{Fundamental Approaches to Software Engineering - 20th
  International Conference, {FASE} 2017, Held as Part of the European Joint
  Conferences on Theory and Practice of Software, {ETAPS} 2017}}, {\sl
  \bibinfo{series}{Lecture Notes in Computer Science}} \bibinfo{volume}{10202},
  \bibinfo{publisher}{Springer}, pp. \bibinfo{pages}{424--441},
  \doi{10.1007/978-3-662-54494-5\_25}.

\bibitemdeclare{inproceedings}{keyabs}
\bibitem{keyabs}
\bibinfo{author}{Crystal~Chang \surnamestart Din\surnameend},
  \bibinfo{author}{Richard \surnamestart Bubel\surnameend} \&
  \bibinfo{author}{Reiner \surnamestart H{\"{a}}hnle\surnameend}
  (\bibinfo{year}{2015}): \emph{\bibinfo{title}{{KeY}-{ABS}: {A} Deductive
  Verification Tool for the Concurrent Modelling Language {ABS}}}.
\newblock In \bibinfo{editor}{Amy~P. \surnamestart Felty\surnameend} \&
  \bibinfo{editor}{Aart \surnamestart Middeldorp\surnameend}, editors: {\sl
  \bibinfo{booktitle}{Intl.\ Conference on Automated Deduction}}, {\sl
  \bibinfo{series}{LNCS}} \bibinfo{volume}{9195},
  \bibinfo{publisher}{Springer}, pp. \bibinfo{pages}{517--526},
  \doi{10.1007/978-3-319-21401-6\_35}.

\bibitemdeclare{article}{ding}
\bibitem{ding}
\bibinfo{author}{Crystal~Chang \surnamestart Din\surnameend} \&
  \bibinfo{author}{Olaf \surnamestart Owe\surnameend} (\bibinfo{year}{2015}):
  \emph{\bibinfo{title}{Compositional reasoning about active objects with
  shared futures}}.
\newblock {\sl \bibinfo{journal}{Formal Aspects of Computing}}
  \bibinfo{volume}{27}(\bibinfo{number}{3}), pp. \bibinfo{pages}{551--572},
  \doi{10.1007/s00165-014-0322-y}.

\bibitemdeclare{inproceedings}{noc}
\bibitem{noc}
\bibinfo{author}{Crystal~Chang \surnamestart Din\surnameend},
  \bibinfo{author}{Silvia Lizeth~Tapia \surnamestart Tarifa\surnameend},
  \bibinfo{author}{Reiner \surnamestart H{\"{a}}hnle\surnameend} \&
  \bibinfo{author}{Einar~Broch \surnamestart Johnsen\surnameend}
  (\bibinfo{year}{2015}): \emph{\bibinfo{title}{History-Based Specification and
  Verification of Scalable Concurrent and Distributed Systems}}.
\newblock In \bibinfo{editor}{Michael~J. \surnamestart Butler\surnameend},
  \bibinfo{editor}{Sylvain \surnamestart Conchon\surnameend} \&
  \bibinfo{editor}{Fatiha \surnamestart Za{\"{\i}}di\surnameend}, editors: {\sl
  \bibinfo{booktitle}{17th International Conference on Formal Engineering
  Methods, {ICFEM} 2015, Proceedings}}, {\sl \bibinfo{series}{Lecture Notes in
  Computer Science}} \bibinfo{volume}{9407}, \bibinfo{publisher}{Springer}, pp.
  \bibinfo{pages}{217--233}, \doi{10.1007/978-3-319-25423-4\_14}.

\bibitemdeclare{inproceedings}{dead}
\bibitem{dead}
\bibinfo{author}{Antonio \surnamestart Flores{-}Montoya\surnameend},
  \bibinfo{author}{Elvira \surnamestart Albert\surnameend} \&
  \bibinfo{author}{Samir \surnamestart Genaim\surnameend}
  (\bibinfo{year}{2013}): \emph{\bibinfo{title}{May-Happen-in-Parallel Based
  Deadlock Analysis for Concurrent Objects}}.
\newblock In: {\sl \bibinfo{booktitle}{Formal Techniques for Distributed
  Systems, {FMOODS/FORTE}}}, pp. \bibinfo{pages}{273--288},
  \doi{10.1007/978-3-642-38592-6\_19}.

\bibitemdeclare{inproceedings}{tut}
\bibitem{tut}
\bibinfo{author}{Reiner \surnamestart H{\"{a}}hnle\surnameend}
  (\bibinfo{year}{2012}): \emph{\bibinfo{title}{The Abstract Behavioral
  Specification Language: {A} Tutorial Introduction}}.
\newblock In \bibinfo{editor}{Elena \surnamestart Giachino\surnameend},
  \bibinfo{editor}{Reiner \surnamestart H{\"{a}}hnle\surnameend},
  \bibinfo{editor}{Frank~S. \surnamestart de~Boer\surnameend} \&
  \bibinfo{editor}{Marcello~M. \surnamestart Bonsangue\surnameend}, editors:
  {\sl \bibinfo{booktitle}{Formal Methods for Components and Objects, 11th
  Intl.\ Symp., {FMCO}, Bertinoro, Italy}}, pp. \bibinfo{pages}{1--37},
  \doi{10.1007/978-3-642-40615-7\_1}.

\bibitemdeclare{inproceedings}{actor}
\bibitem{actor}
\bibinfo{author}{Carl \surnamestart Hewitt\surnameend}, \bibinfo{author}{Peter
  \surnamestart Bishop\surnameend} \& \bibinfo{author}{Richard \surnamestart
  Steiger\surnameend} (\bibinfo{year}{1973}): \emph{\bibinfo{title}{A universal
  modular {ACTOR} formalism for artificial intelligence}}.
\newblock In: {\sl \bibinfo{booktitle}{Proceedings of the 3rd International
  Joint Conference on Artificial Intelligence}}, \bibinfo{series}{IJCAI'73},
  \bibinfo{publisher}{Morgan Kaufmann Publishers Inc.}, pp.
  \bibinfo{pages}{235--245}.
\newblock \urlprefix\url{http://dl.acm.org/citation.cfm?id=1624775.1624804}.

\bibitemdeclare{book}{promela}
\bibitem{promela}
\bibinfo{author}{Gerard~J. \surnamestart Holzmann\surnameend}
  (\bibinfo{year}{1991}): \emph{\bibinfo{title}{Design and Validation of
  Computer Protocols}}.
\newblock \bibinfo{publisher}{Prentice-Hall, Inc.}

\bibitemdeclare{inproceedings}{ABS}
\bibitem{ABS}
\bibinfo{author}{Einar~Broch \surnamestart Johnsen\surnameend},
  \bibinfo{author}{Reiner \surnamestart H{\"{a}}hnle\surnameend},
  \bibinfo{author}{Jan \surnamestart Sch{\"{a}}fer\surnameend},
  \bibinfo{author}{Rudolf \surnamestart Schlatte\surnameend} \&
  \bibinfo{author}{Martin \surnamestart Steffen\surnameend}
  (\bibinfo{year}{2010}): \emph{\bibinfo{title}{{ABS:} {A} Core Language for
  Abstract Behavioral Specification}}.
\newblock In \bibinfo{editor}{Bernhard~K. \surnamestart Aichernig\surnameend},
  \bibinfo{editor}{Frank~S. \surnamestart de~Boer\surnameend} \&
  \bibinfo{editor}{Marcello~M. \surnamestart Bonsangue\surnameend}, editors:
  {\sl \bibinfo{booktitle}{Formal Methods for Components and Objects, 9th
  Intl.\ Symp., {FMCO}}}, pp. \bibinfo{pages}{142--164},
  \doi{10.1007/978-3-642-25271-6\_8}.

\bibitemdeclare{inproceedings}{avocs}
\bibitem{avocs}
\bibinfo{author}{Eduard \surnamestart Kamburjan\surnameend}
  (\bibinfo{year}{2018}): \emph{\bibinfo{title}{Detecting Deadlocks in Formal
  System Models with Condition Synchronization}}.
\newblock In: {\sl \bibinfo{booktitle}{Accepted for Publication at AVoCS'18}}.

\bibitemdeclare{inproceedings}{ftscs}
\bibitem{ftscs}
\bibinfo{author}{Eduard \surnamestart Kamburjan\surnameend} \&
  \bibinfo{author}{Reiner \surnamestart H{\"a}hnle\surnameend}
  (\bibinfo{year}{2017}): \emph{\bibinfo{title}{Uniform Modeling of Railway
  Operations}}.
\newblock In \bibinfo{editor}{Cyrille \surnamestart Artho\surnameend} \&
  \bibinfo{editor}{Peter~Csaba \surnamestart {\"O}lveczky\surnameend}, editors:
  {\sl \bibinfo{booktitle}{Formal Techniques for Safety-Critical Systems: 5th
  Intl.\ Workshop, FTSCS, Revised Selected Papers}},
  \bibinfo{publisher}{Springer}, pp. \bibinfo{pages}{55--71},
  \doi{10.1007/978-3-319-53946-1\_4}.

\bibitemdeclare{inproceedings}{cyber1}
\bibitem{cyber1}
\bibinfo{author}{Ehsan \surnamestart Khamespanah\surnameend},
  \bibinfo{author}{Kirill \surnamestart Mechitov\surnameend},
  \bibinfo{author}{Marjan \surnamestart Sirjani\surnameend} \&
  \bibinfo{author}{Gul~A. \surnamestart Agha\surnameend}
  (\bibinfo{year}{2016}): \emph{\bibinfo{title}{Schedulability Analysis of
  Distributed Real-Time Sensor Network Applications Using Actor-Based Model
  Checking}}.
\newblock In \bibinfo{editor}{Dragan \surnamestart Bosnacki\surnameend} \&
  \bibinfo{editor}{Anton \surnamestart Wijs\surnameend}, editors: {\sl
  \bibinfo{booktitle}{Model Checking Software, 23rd Intl.\ Symp., {SPIN}}}, pp.
  \bibinfo{pages}{165--181}, \doi{10.1007/978-3-319-32582-8\_11}.

\bibitemdeclare{article}{lamport}
\bibitem{lamport}
\bibinfo{author}{Leslie \surnamestart Lamport\surnameend}
  (\bibinfo{year}{1979}): \emph{\bibinfo{title}{How to Make a Multiprocessor
  Computer That Correctly Executes Multiprocess Programs}}.
\newblock {\sl \bibinfo{journal}{{IEEE} Trans. Computers}}
  \bibinfo{volume}{28}(\bibinfo{number}{9}), pp. \bibinfo{pages}{690--691},
  \doi{10.1109/TC.1979.1675439}.

\bibitemdeclare{inproceedings}{hadoop}
\bibitem{hadoop}
\bibinfo{author}{Jia{-}Chun \surnamestart Lin\surnameend},
  \bibinfo{author}{Ingrid~Chieh \surnamestart Yu\surnameend},
  \bibinfo{author}{Einar~Broch \surnamestart Johnsen\surnameend} \&
  \bibinfo{author}{Ming{-}Chang \surnamestart Lee\surnameend}
  (\bibinfo{year}{2016}): \emph{\bibinfo{title}{{ABS-YARN:} {A} Formal
  Framework for Modeling Hadoop {YARN} Clusters}}.
\newblock In \bibinfo{editor}{Perdita \surnamestart Stevens\surnameend} \&
  \bibinfo{editor}{Andrzej \surnamestart Wasowski\surnameend}, editors: {\sl
  \bibinfo{booktitle}{Fundamental Approaches to Software Engineering, 19th
  Intl.\ Conf., {FASE}}}, pp. \bibinfo{pages}{49--65},
  \doi{10.1007/978-3-662-49665-7\_4}.

\bibitemdeclare{inproceedings}{Mantel}
\bibitem{Mantel}
\bibinfo{author}{Heiko \surnamestart Mantel\surnameend},
  \bibinfo{author}{Matthias \surnamestart Perner\surnameend} \&
  \bibinfo{author}{Jens \surnamestart Sauer\surnameend} (\bibinfo{year}{2014}):
  \emph{\bibinfo{title}{Noninterference under Weak Memory Models}}.
\newblock In: {\sl \bibinfo{booktitle}{{IEEE} 27th Computer Security
  Foundations Symp., {CSF}}}, \bibinfo{publisher}{{IEEE} Computer Society}, pp.
  \bibinfo{pages}{80--94}, \doi{10.1109/CSF.2014.14}.

\bibitemdeclare{inproceedings}{absspl}
\bibitem{absspl}
\bibinfo{author}{Radu \surnamestart Muschevici\surnameend},
  \bibinfo{author}{Dave \surnamestart Clarke\surnameend} \&
  \bibinfo{author}{Jos{\'{e}} \surnamestart Proen{\c{c}}a\surnameend}
  (\bibinfo{year}{2013}): \emph{\bibinfo{title}{Executable modelling of dynamic
  software product lines in the {ABS} language}}.
\newblock In \bibinfo{editor}{Andreas \surnamestart Classen\surnameend} \&
  \bibinfo{editor}{Norbert \surnamestart Siegmund\surnameend}, editors: {\sl
  \bibinfo{booktitle}{5th Intl.\ Workshop on Feature-Oriented Software
  Development, {FOSD}}}, pp. \bibinfo{pages}{17--24},
  \doi{10.1145/2528265.2528266}.

\bibitemdeclare{inproceedings}{cafe}
\bibitem{cafe}
\bibinfo{author}{Shin \surnamestart Nakajima\surnameend} \&
  \bibinfo{author}{Kokichi \surnamestart Futatsugi\surnameend}
  (\bibinfo{year}{1997}): \emph{\bibinfo{title}{An Object-Oriented Modeling
  Method for Algebraic Specifications in {CafeOBJ}}}.
\newblock In \bibinfo{editor}{W.~Richards \surnamestart Adrion\surnameend},
  \bibinfo{editor}{Alfonso \surnamestart Fuggetta\surnameend},
  \bibinfo{editor}{Richard~N. \surnamestart Taylor\surnameend} \&
  \bibinfo{editor}{Anthony~I. \surnamestart Wasserman\surnameend}, editors:
  {\sl \bibinfo{booktitle}{Pulling Together, Proc.\ 19th Int.\ Conf.\ on
  Software Engineering}}, pp. \bibinfo{pages}{34--44},
  \doi{10.1145/253228.253238}.

\bibitemdeclare{book}{isabelle}
\bibitem{isabelle}
\bibinfo{author}{Tobias \surnamestart Nipkow\surnameend},
  \bibinfo{author}{Markus \surnamestart Wenzel\surnameend} \&
  \bibinfo{author}{Lawrence~C. \surnamestart Paulson\surnameend}
  (\bibinfo{year}{2002}): \emph{\bibinfo{title}{Isabelle/HOL: A Proof Assistant
  for Higher-order Logic}}.
\newblock \bibinfo{publisher}{Springer}, \doi{10.1007/3-540-45949-9}.

\bibitemdeclare{book}{splbook}
\bibitem{splbook}
\bibinfo{author}{Klaus \surnamestart Pohl\surnameend},
  \bibinfo{author}{G{\"{u}}nter \surnamestart B{\"{o}}ckle\surnameend} \&
  \bibinfo{author}{Frank \surnamestart van~der Linden\surnameend}
  (\bibinfo{year}{2005}): \emph{\bibinfo{title}{Software Product Line
  Engineering - Foundations, Principles, and Techniques}}.
\newblock \bibinfo{publisher}{Springer}, \doi{10.1007/3-540-28901-1}.

\bibitemdeclare{book}{variable}
\bibitem{variable}
\bibinfo{author}{Klaus \surnamestart Pohl\surnameend},
  \bibinfo{author}{G\"{u}nter \surnamestart B\"{o}ckle\surnameend} \&
  \bibinfo{author}{Frank J. van~der \surnamestart Linden\surnameend}
  (\bibinfo{year}{2005}): \emph{\bibinfo{title}{Software Product Line
  Engineering: Foundations, Principles and Techniques}}.
\newblock \bibinfo{publisher}{Springer-Verlag New York, Inc.},
  \doi{10.1007/3-540-28901-1}.

\bibitemdeclare{inproceedings}{Axiom2}
\bibitem{Axiom2}
\bibinfo{author}{Vijay~A. \surnamestart Saraswat\surnameend},
  \bibinfo{author}{Radha \surnamestart Jagadeesan\surnameend},
  \bibinfo{author}{Maged \surnamestart Michael\surnameend} \&
  \bibinfo{author}{Christoph \surnamestart von Praun\surnameend}
  (\bibinfo{year}{2007}): \emph{\bibinfo{title}{A Theory of Memory Models}}.
\newblock In: {\sl \bibinfo{booktitle}{Proceedings of the 12th ACM SIGPLAN
  Symposium on Principles and Practice of Parallel Programming}},
  \bibinfo{series}{PPoPP '07}, \bibinfo{publisher}{ACM}, pp.
  \bibinfo{pages}{161--172}, \doi{10.1145/1229428.1229469}.

\bibitemdeclare{inproceedings}{Sarkar}
\bibitem{Sarkar}
\bibinfo{author}{Susmit \surnamestart Sarkar\surnameend},
  \bibinfo{author}{Peter \surnamestart Sewell\surnameend},
  \bibinfo{author}{Jade \surnamestart Alglave\surnameend}, \bibinfo{author}{Luc
  \surnamestart Maranget\surnameend} \& \bibinfo{author}{Derek \surnamestart
  Williams\surnameend} (\bibinfo{year}{2011}):
  \emph{\bibinfo{title}{Understanding {POWER} multiprocessors}}.
\newblock In \bibinfo{editor}{Mary~W. \surnamestart Hall\surnameend} \&
  \bibinfo{editor}{David~A. \surnamestart Padua\surnameend}, editors: {\sl
  \bibinfo{booktitle}{Proceedings of the 32nd {ACM} {SIGPLAN} Conference on
  Programming Language Design and Implementation, {PLDI} 2011}},
  \bibinfo{publisher}{{ACM}}, pp. \bibinfo{pages}{175--186},
  \doi{10.1145/1993498.1993520}.

\bibitemdeclare{article}{Schaefer2012}
\bibitem{Schaefer2012}
\bibinfo{author}{Ina \surnamestart Schaefer\surnameend}, \bibinfo{author}{Rick
  \surnamestart Rabiser\surnameend}, \bibinfo{author}{Dave \surnamestart
  Clarke\surnameend}, \bibinfo{author}{Lorenzo \surnamestart
  Bettini\surnameend}, \bibinfo{author}{David \surnamestart
  Benavides\surnameend}, \bibinfo{author}{Goetz \surnamestart
  Botterweck\surnameend}, \bibinfo{author}{Animesh \surnamestart
  Pathak\surnameend}, \bibinfo{author}{Salvador \surnamestart
  Trujillo\surnameend} \& \bibinfo{author}{Karina \surnamestart
  Villela\surnameend} (\bibinfo{year}{2012}): \emph{\bibinfo{title}{Software
  diversity: state of the art and perspectives}}.
\newblock {\sl \bibinfo{journal}{International Journal on Software Tools for
  Technology Transfer}} \bibinfo{volume}{14}(\bibinfo{number}{5}), pp.
  \bibinfo{pages}{477--495}, \doi{10.1007/s10009-012-0253-y}.

\bibitemdeclare{inbook}{kmaude}
\bibitem{kmaude}
\bibinfo{author}{Traian~Florin \surnamestart
  {\c{S}}erb{\u{a}}nu{\c{t}}{\u{a}}\surnameend} \& \bibinfo{author}{Grigore
  \surnamestart Ro{\c{s}}u\surnameend} (\bibinfo{year}{2010}):
  \emph{\bibinfo{title}{K-Maude: A Rewriting Based Tool for Semantics of
  Programming Languages}}, pp. \bibinfo{pages}{104--122}.
\newblock \bibinfo{publisher}{Springer Berlin Heidelberg},
  \doi{10.1007/978-3-642-16310-4\_8}.

\bibitemdeclare{article}{xtso}
\bibitem{xtso}
\bibinfo{author}{Peter \surnamestart Sewell\surnameend},
  \bibinfo{author}{Susmit \surnamestart Sarkar\surnameend},
  \bibinfo{author}{Scott \surnamestart Owens\surnameend},
  \bibinfo{author}{Francesco~Zappa \surnamestart Nardelli\surnameend} \&
  \bibinfo{author}{Magnus~O. \surnamestart Myreen\surnameend}
  (\bibinfo{year}{2010}): \emph{\bibinfo{title}{{X86-TSO}: A Rigorous and
  Usable Programmer's Model for x86 Multiprocessors}}.
\newblock {\sl \bibinfo{journal}{Commun. ACM}}
  \bibinfo{volume}{53}(\bibinfo{number}{7}), pp. \bibinfo{pages}{89--97},
  \doi{10.1145/1785414.1785443}.

\bibitemdeclare{article}{noc2}
\bibitem{noc2}
\bibinfo{author}{Zeinab \surnamestart Sharifi\surnameend},
  \bibinfo{author}{Mahdi \surnamestart Mosaffa\surnameend},
  \bibinfo{author}{Siamak \surnamestart Mohammadi\surnameend} \&
  \bibinfo{author}{Marjan \surnamestart Sirjani\surnameend}
  (\bibinfo{year}{2013}): \emph{\bibinfo{title}{Functional and Performance
  Analysis of Network-on-Chips Using Actor-based Modeling and Formal
  Verification}}.
\newblock {\sl \bibinfo{journal}{{ECEASST}}} \bibinfo{volume}{66}.
\newblock \urlprefix\url{http://dx.doi.org/10.14279/tuj.eceasst.66.890}.

\bibitemdeclare{article}{rebeca}
\bibitem{rebeca}
\bibinfo{author}{Marjan \surnamestart Sirjani\surnameend}, \bibinfo{author}{Ali
  \surnamestart Movaghar\surnameend}, \bibinfo{author}{Amin \surnamestart
  Shali\surnameend} \& \bibinfo{author}{Frank~S. \surnamestart
  de~Boer\surnameend} (\bibinfo{year}{2004}): \emph{\bibinfo{title}{Modeling
  and Verification of Reactive Systems using Rebeca}}.
\newblock {\sl \bibinfo{journal}{Fundam. Inform.}}
  \bibinfo{volume}{63}(\bibinfo{number}{4}), pp. \bibinfo{pages}{385--410}.
\newblock
  \urlprefix\url{http://content.iospress.com/articles/fundamenta-informaticae/fi63-4-05}.

\bibitemdeclare{manual}{coq}
\bibitem{coq}
\bibinfo{author}{\surnamestart {The Coq development team}\surnameend}
  (\bibinfo{year}{2004}): \emph{\bibinfo{title}{The Coq proof assistant
  reference manual}}.
\newblock \bibinfo{organization}{LogiCal Project}.
\newblock \urlprefix\url{http://coq.inria.fr}.
\newblock \bibinfo{note}{Version 8.0}.

\bibitemdeclare{mastersthesis}{weber}
\bibitem{weber}
\bibinfo{author}{Alexandra \surnamestart Weber\surnameend}
  (\bibinfo{year}{2014}): \emph{\bibinfo{title}{Comparison of an operational
  and an axiomatic model of execution for multi-threaded programs}}.
\newblock Master's thesis, \bibinfo{school}{{TU Darmstadt}}.
\newblock \urlprefix\url{https://hds.hebis.de/ulbda/Record/HEB350072914}.

\end{thebibliography}
\end{document}